\documentclass[
  prl,
  twocolumn,
  showpacs,
  preprintnumbers,
  amsmath,
  amssymb
]{revtex4}
\usepackage{epsf}
\usepackage{amsfonts}
\usepackage{amssymb}
\usepackage{amsthm}
\usepackage{graphicx}
\def\be{\begin{equation}}
\def\ee{\end{equation}}
\def\bea{\begin{eqnarray}}
\def\eea{\end{eqnarray}}
\def\ua{\uparrow}
\def\da{\downarrow}

\def\witchbox#1#2#3{\hbox{$\mathchar"#1#2#3$}}
\def\leqsim{\mathrel{\rlap{\lower3pt\witchbox218}\raise2pt\witchbox13C}}
\def\geqsim{\mathrel{\rlap{\lower3pt\witchbox218}\raise2pt\witchbox13E}}
\begin{document}

%\date{\today}

\title{Quantitative determination of the Hubbard model phase 
diagram from optical lattice experiments by two-parameter scaling}

\author{V. L. Campo, Jr.$^1$, K. Capelle$^2$, J. Quintanilla$^{3,*}$, and C. Hooley$^4$}

\affiliation{$\mbox{}^1$Centro Internacional de F{\'\i}sica de Mat\'eria Condensada, Universidade de Bras{\'\i}lia, Caixa Postal 04513, 70919-970 Bras{\'\i}lia, Brazil}

\affiliation{$\mbox{}^2$Departamento de F{\'\i}sica e Inform{\'a}tica, Instituto de F{\'\i}sica de S{\~a}o Carlos, Universidade de S{\~a}o Paulo, Caixa Postal 369, 13560-970 S{\~a}o Carlos, S{\~a}o Paulo, Brazil}

\affiliation{$\mbox{}^3$ISIS facility, STFC Rutherford Appleton Laboratory, Harwell Science and Innovation Campus, Didcot OX11 0QX, U.K.}

\affiliation{$\mbox{}^4$School of Physics and Astronomy, University of St Andrews, North Haugh, St Andrews, Fife KY16 9SS, U.K.}

\begin{abstract}
We propose an experiment to obtain the phase diagram of the fermionic
Hubbard model, for any dimensionality, using cold atoms in optical
lattices.  It is based on measuring the total energy for a sequence of
trap profiles.  It combines finite-size scaling with an additional
`finite-curvature scaling' necessary to reach the homogeneous limit.  We
illustrate its viability in the 1D case, simulating experimental data in
the Bethe-Ansatz local density approximation.  Including experimental
errors, the filling corresponding to the Mott transition can be
determined with better than 3\% accuracy.
\end{abstract}

%\keywords{word,word,word}

\maketitle

Bose-Einstein condensation of trapped atoms \cite{firstBEC} has led to rapid growth of research into such systems.  One exciting suggestion \cite{simulateCM} is that they could be used to simulate condensed matter, with the atoms playing the role of conduction electrons, and a periodic potential provided by a laser standing wave, or `optical lattice'.  The advantages of such a scheme are manifold: the system is highly tunable, including the effective strength of atom-atom interactions \cite{feshbach}; there is no naturally occurring disorder in the lattice; and quantities such as the total energy, which cannot normally be measured in condensed matter systems, are accessible \cite{totalenergy}.  Several familiar phases of matter have already been seen:\ the Fermi liquid \cite{fermiliquidobs}, the Mott insulator \cite{mottinsulatorobs}, vortex matter \cite{vortexmatterobs}, and BCS condensates \cite{bcscondensateobs}.  So far, however, most observations have been (a) rather qualitative in nature, and (b) of phases that were already well understood.  In this Letter, we propose a scheme to answer \emph{quantitatively} an outstanding \emph{open} question of condensed matter physics: what is the phase diagram of the fermionic Hubbard model in two or three dimensions?

To answer this question we need to address the fact that atom traps are inhomogeneous. Because of that, phase transitions are `blurred' \cite{blurring}. This is not simply an issue of finite size:\ the system is not in a thermodynamic phase at all, even in the thermodynamic limit (see below). Thus, the notion of a phase diagram
for the atomic gas in the optical lattice is not rigorous. Nevertheless, by focusing on the total energy and carrying out a sequence of measurements for traps of different sizes and shapes the phase diagram of the model of interest may be deduced, as we shall show.

To achieve this, we need to link quantitatively the energy of the model system (atoms in an optical lattice) to that of an infinite
$d$-dimensional Hubbard lattice. One obvious issue is size:\ condensed matter systems typically have $N \sim 10^{26}$ conduction electrons, whereas $N \sim 10^5$ is more typical in the optical lattice case.  In computer simulations, this is dealt with by {\it finite-size scaling} \cite{finitesizescaling}:\ systems of various sizes are simulated, and the results extrapolated to the thermodynamic limit.  A similar procedure can be employed in optical lattice experiments, as we describe below.

Another difference between solid-state and cold-atom systems is the nature of the confining potential.  In the former, it is usually a hard-wall box; in the latter, it is a power-law (usually harmonic) potential. The thermodynamic limit mentioned above is \emph{singular} \cite{2004-Hooley-Quintanilla,2006-Hooley-Quintanilla}:\ even though it corresponds to vanishing trapping potential, the results obtained in that limit depend on the shape of the trap. Thus finite-size scaling alone does not yield the desired bulk energy, and a second kind of scaling is called for:\ an extrapolation from the properties of the power-law-trapped system to those of the hard-wall-confined one.  We call this {\it finite-curvature scaling}.

\begin{figure}
\begin{center}
\includegraphics[height=\columnwidth,angle=-90]{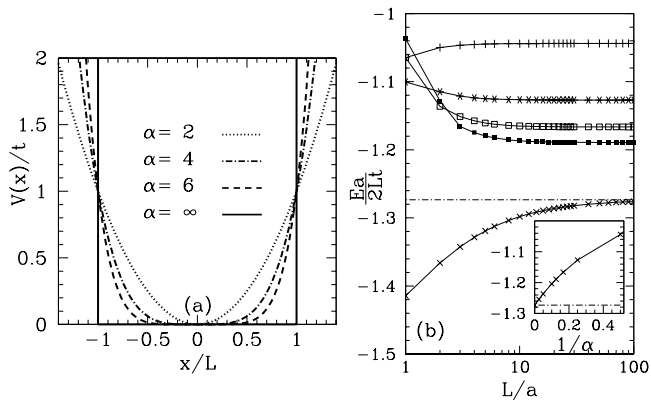}
\end{center}
\caption{(a) The trap potential $V(x)$ for $d=1$ and various values of $\alpha$.  (b) Dependence of total energy on system size for a $d=1$ optical lattice loaded with non-interacting, spin-$1/2$ fermions.  The filling is fixed at $f=1$.  The trap exponents are $\alpha=2$ (+), 4 (*), 6 ($\square$), 8 ($\blacksquare$) and $\infty$ ($\times$).
Inset: the same quantity in the thermodynamic limit plotted against 
$1/\alpha$. The dashed-dotted lines mark the thermodynamic limit result 
for $\alpha=\infty$:  $Ea/2Lt =-(4/\pi)\sin(\pi f/2)$.}
\label{Fig.1}
\end{figure}

Thus we consider the family of potentials given by \cite{2006-Hooley-Quintanilla}
$V({\bf x}) = t \sum_{i=1}^d \left\vert x_i/L_i \right\vert^\alpha$,
where $L_i$ are characteristic length-scales of the trap, and $t$ is a reference energy scale (see below).  $\alpha = 2$ corresponds to harmonic trapping, while the $\alpha \to \infty$ limit creates a hard-wall box of volume
$2^d \prod_i L_i$. The $d=1$ version of this potential is shown in 
Fig.~\ref{Fig.1}(a).

In practice, the achievable values of $\alpha$ are likely to be the even 
integers, because the true confining potential will generically be 
harmonic ($\sim x^2$) near its minimum.  To change this one 
will have to superpose two laser beams to cancel the harmonic terms; for Gaussian beams, the next term will be quartic ($\sim 
x^4$).  Additional laser beams could be used to cancel both the $x^2$
and 
$x^4$ terms, leaving an $x^6$ potential, and so on. This demands controlling the width of each Gaussian beam independently of its wavelength, which is within current experimental capabilities \cite{knots,2006-Boyer-et-al}. Indeed, the realisation of a confining potential with a term $V(x) \sim x^4$ has already been reported \cite{quarticpotential}.

The experiments we propose are realisations of the following inhomogeneous Hubbard model \cite{simulateCM,OLcalculations}:
\begin{eqnarray}
H & = & -t \sum_{\langle ij \rangle \sigma} \left( c^\dagger_{i\sigma} c_{j\sigma} + {\rm H.c.} \right)
+ U \sum_j c^\dagger_{j\ua} c_{j\ua} c^\dagger_{j\da} c_{j\da} \nonumber \\
& & \qquad + \sum_{j\sigma} V({\bf x}_j) c^\dagger_{j\sigma} c_{j\sigma}, \label{ham}
\end{eqnarray}
where $c^\dagger_{j\sigma}$ creates a fermionic atom with spin $\sigma$ on site $j$.
The terms in (\ref{ham}) represent respectively the kinetic energy, the potential energy of atom-atom interactions, and the potential energy due to the trap.  The usual Hubbard model, whose phase diagram we wish to deduce, is recovered when $V({\bf x}_j)=0$ everywhere. The atoms must have two `spin' states ($\ua$ and $\da$); in practice these will be two hyperfine eigenstates.

The model (\ref{ham}) has four dimensionless parameters: $U/t$, the strength of the on-site repulsion; $\alpha$, the trap exponent; $\mathcal{N}$, the number of sites; and $f$, the filling.  These last two are somewhat delicate, since in the presence of a power-law trap the effective number of sites in the lattice becomes energy-dependent.  We thus define $\mathcal{N}$ as the number of sites {\it in the} $\alpha \to \infty$ {\it limit}, i.e.\ $\mathcal{N} = \prod_{i=1}^d (2L_i/a)$, and $f$ as $N/\mathcal{N}$.
The thermodynamic limit is
\be
 \mathcal{N}\to\infty, \,N \to \infty, \,\mbox{with}\,f= {\rm constant}
\label{thermlim}
\ee
for given $U/t$ and $\alpha$, coinciding with the usual definition for $\alpha \to \infty$ and with Ref.~\cite{1996-Damle-Senthil-Majumdar-Sachdev} for $\alpha=2$. In the limit (\ref{thermlim}), the trapping potential term in (\ref{ham}) vanishes identically.

To make the case for finite-curvature in addition to finite-size scaling, 
we consider first the non-interacting situation. For large enough $\mathcal{N}$, the density of states (DOS) obeys the scaling law 
\(
  \rho\left(\epsilon\right) 
  = 
  \frac{\mathcal N}{t}G
    \left(
      \frac{\epsilon}{t}
    \right).
\)
The derivation of $G(x)$, which depends implicitly on $d$ and $\alpha$, has been outlined in Refs.~\cite{2004-Hooley-Quintanilla,2006-Hooley-Quintanilla} and its behaviour for $x \gg 1$ has been verified experimentally in the 1D case for $\alpha=2$ \cite{2004-Ott-et-al}. If we define 
$F_{\alpha}\left(\frac{\mu}{t}\right) = \int_{-2}^{\mu/t} dx
G(x)x,$ and fix the Lagrange multiplier $\mu$ by
%%$f=\int_0^{\mu/t} dxF(x),$
$f=\int_{-2}^{\mu/t} dxG(x),$
the ground-state energy obeys the scaling law
\be
  E = t \mathcal{N} F_\alpha\left(f\right).
  \label{fss1}
\ee
Thus $F_\alpha(f)$ is the energy per site in the thermodynamic limit (\ref{thermlim}) in units of $t$. This allows its determination from results for finite size systems by extrapolation.

Fig.~\ref{Fig.1}(b) shows some $d=1$ total energy curves.  The energy per site has converged essentially perfectly for system sizes $\mathcal{N} \sim 200$ or larger. This makes it possible to obtain $F_\alpha(x)$ from experiments on reasonably small systems. 
However, the scaling function $F_\alpha(x)$ itself is {\em not} universal, but depends on the trap exponent, $\alpha$ (see figure). An additional extrapolation in $\alpha$ is therefore necessary to obtain the $\alpha \to \infty$ behaviour (see inset).
\begin{figure}
\begin{center}
  \includegraphics[height=\columnwidth,angle=-90]{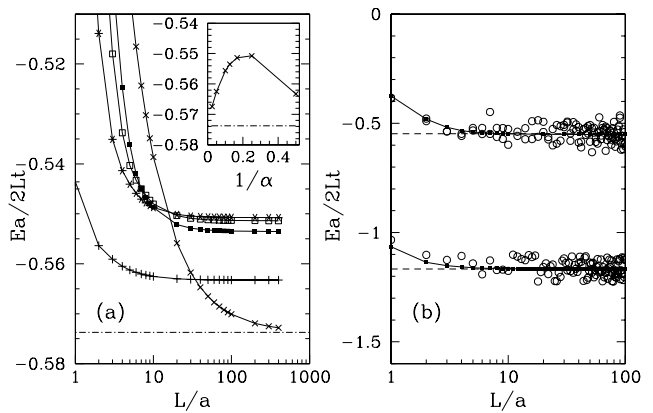}
\end{center}
  \caption{
(a) Same as Fig.~\ref{Fig.1}(b), but with interaction strength $U/t=4$, 
and with $L/a=400$ data used for the inset.  In this case the 
thermodynamic limit result for $\alpha=\infty$ is $E=-0.57372937$.  (b) 
Solid data points and line: the $\alpha=6$ data from 
Figs.~\ref{Fig.1}(b) (lower curve) and \ref{Fig.2}(a) (upper curve).  Open data points: the same, but with random errors of $\pm 5\%$ added to the values of $E$ and $f$.  Dashed line: least-squares fit to these `noisy' data.  The resulting error in the thermodynamic limit of the energy is $\leqsim 0.6\%$.}
\label{Fig.2}
\end{figure}

We now turn to the case of interest, $U/t > 0$. For $\alpha \to \infty$ and in the thermodynamic limit (\ref{thermlim}), the model (\ref{ham}) is known to exhibit at least two distinct phases:\ a metallic one at low or high filling, and an insulating one at (and perhaps also around) $f=1$.  The 1D Hubbard model has been solved exactly \cite{1968-Lieb-Wu}:\ it is metallic except at precisely $f=1$, where it is insulating for all $U>0$.  But the behaviour of the $d>1$ models remains uncertain; for example, it is not clear whether the insulating phase occupies just the line $f=1$, or some finite region of the phase diagram. Moreover, there may be additional competing phases, including $d$-wave superconductivity, itinerant antiferromagnetism, and phases with a distorted Fermi surface \cite{phases2dhub}.

Our proposal for investigating this experimentally is as follows: (a) Obtain experimental values for the total energy for several values of the filling fraction, trap power-law, system size, and interatomic interaction strength. 
This builds up data points from the function $E(f,U,t,\alpha,\mathcal{N})$. 
(b) Obtain the values of $E$ in the thermodynamic limit (\ref{thermlim}) by extrapolation, aided by the finite-size scaling law (\ref{fss1}), where $F_\alpha(f)$ now additionally depends on $U/t$. (c) Numerically differentiate the curves thus obtained with respect to the filling $f$, to obtain the chemical potential $\mu(f,U,t,\alpha) = tF_\alpha'(f,U/t)$. (d) Take the $\alpha \to \infty$ limit by a second extrapolation. This is the finite-curvature scaling, and results in a function $\mu/t$ of $f$ and $U/t$. (e) By finding the lines in the $f-U/t$ plane where $\mu/t$ is discontinuous, construct the phase diagram.

To demonstrate our method in the one-dimensional case, where the answer is already known \cite{1968-Lieb-Wu}, we have used density-functional theory within the Bethe-Ansatz local-density approximation (BA-LDA) \cite{BA-LDA} to simulate the experimental data for step (a).
The BA-LDA is known to predict ground-state energies of spatially 
inhomogeneous systems with an accuracy of a few percent.  Furthermore, 
its accuracy increases as the length-scale of the inhomogeneity 
increases, making it particularly suitable to study the approach to the 
thermodynamic limit (\ref{thermlim}) \cite{footnote3}.

Examples of the resulting total energy curves are shown in Fig.~\ref{Fig.2}(a) for $U/t=4$. Note that a scaling law of the form (\ref{fss1}) is still obeyed;
indeed, (\ref{fss1}) can be obtained using only the requirement that, for large $\mathcal{N}$, $E$ is extensive. We find that to achieve good convergence $\mathcal{N} \sim 200$ sites suffice for $\alpha \leq 8$.  Experimentally, this would require trap frequencies $\sim 10\,{\rm Hz}$ in the $^6$Li case, which is in the currently achievable range. Similarly fast convergence is found for larger $U/t=6,8$ (not shown).

This rapid convergence has the important effect of rendering the method rather insensitive to experimental error.  For example, we show in Fig.~\ref{Fig.2}(b) the $\alpha=6$ curves from Figs.~\ref{Fig.1}(b) and \ref{Fig.2}(a), but with random errors of $\pm 5\%$ added to the values of $f$ and $E$.  Though these data look very noisy, all points for $L/a \geqsim 30$ are essentially random scatter around the same value; thus the central limit theorem applies, and the overall error may be reduced by increasing the number of measurements.  The dashed lines are least-squares fits to the noisy data for $30 \leqslant L/a \leqslant 100$; their differences from the noise-free values are 0.17\% (non-interacting case) and 0.61\% (interacting case).  Thus a serious error (cumulatively $\sim \pm 8\%$ or worse, as Fig.~\ref{Fig.2}(b) shows) in the individual data points becomes a negligible one ($\leqsim 0.6\%$) in the resulting value for the thermodynamic limit of the energy.

\begin{figure}
\begin{center}
 \includegraphics[height=\columnwidth,angle=-90]{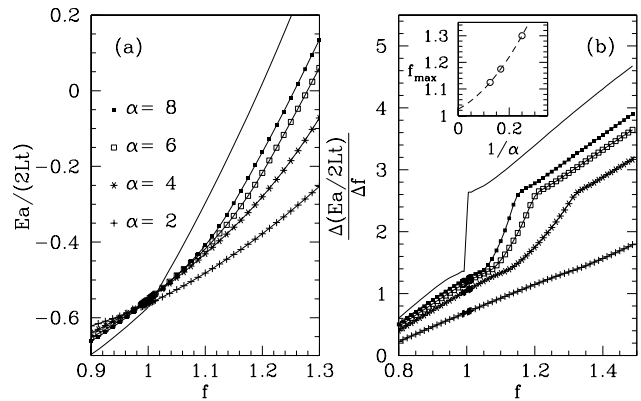}
\end{center}
\caption{Energy per site (a) and numerical estimate of the chemical potential (b) vs $f$ for $U/t=4$, $L/a=400$, $\alpha=2$, $4$, $6$, $8$ and $\infty$ (lines with no symbols).
In the inset in panel (b), the filling $f_{max}$ where $\mu(f)$ has maximum slope is plotted as a function of $\alpha^{-1}$ (circles).  The parabola interpolating
the points coming from $\alpha = 4$, $6$ and $8$ (dashed line) extrapolates to $f_{max}(0) = 
1.025$, indicating that even those relatively small trap exponents
provide a good estimate of the critical filling.}
\label{Fig.3}
\end{figure}

In Fig.~\ref{Fig.3}(a), we plot energy per site as a function of $f$ for $\alpha=2,4,6,8,\infty$ and system size ${\mathcal N}=800$ (effectively in the thermodynamic limit). This must be intepreted as a plot of the extrapolated values of $E$ vs nominal values of $f$. The error discussed above (coming from the uncertainty in measuring each individual value of $E$, on the one hand, and the fact that the actual filling differs from the nominal one for each particular measurement, on the other) is smaller than the size of the symbols. The cusp indicating the phase transition becomes sharper as $\alpha$ is increased, and tends towards its hard-wall position 
at $f=1$ as $\alpha \to \infty$, but for finite $\alpha$ it is not a singularity. Numerically differentiating these curves with respect to $f$, shown in Fig.~\ref{Fig.3}(b), makes the cusp more apparent and shows that there is a single phase transition at $f \approx 1$. The inset demonstrates a simple method that allows us to obtain the critical value of $f$ within $\sim 3\%$ (or $\sim 7\%$ if only the data points with $\alpha \leq 6$ are used). Clearly, the same method can also be applied in $d=2$ and $d=3$, where the exact answer is not known. 

The implementation of our proposal requires a significant degree of automation.  However, preparation-to-measurement times of less than a minute are already routinely achieved, and complete automation is being developed for drop-tower-based zero-gravity experiments \cite{zerograv}, so this requirement is not unfeasible. One could alternatively avoid the finite-curvature issues by creating a `flat' potential, using the endcap-beam technique \cite{endcap}.  However, Fig.~\ref{Fig.2}(a) shows that the convergence with system-size is much slower in this case, which would lead to appreciably larger errors.

In conclusion, we have presented an experimental scheme to determine the Hubbard model phase diagram by measuring the total energy of optical lattice systems. It requires a non-trivial curvature extrapolation in addition to the usual finite-size one. We have demonstrated its validity in the $d=1$ case by an approximate numerical many-body 
calculation to simulate experimental data, including an analysis of the effects of realistic experimental errors. We conclude that the fermionic Hubbard model's phase diagram can be determined quantitatively using the experimental procedure that we propose.

We stress that the experimental use of our method would
yield the location of the phase boundaries of the Hubbard model as functions of interaction
strength, $U$, and filling, $f$. This would be a significant step forward 
in $d=2$ and $d=3$, where standard computational
approaches encounter serious difficulties and relatively little is known for certain about the phase diagram.
It would not, however,
permit the identification of the phases; for this, further experiments
would be required.  An obvious choice would be to probe correlation functions, e.g.\ {\it via} the dynamical structure factor $S({\bf q},\omega)$ using stimulated emission in a two-laser set-up \cite{structurefactor} or shot-noise techniques \cite{shotnoise}. Alternatively, one could exploit the fluctuation-dissipation theorem to derive them from the total energy; we will discuss this elsewhere.

We thank P. W. Courteille, M. R. Dennis, M. Foulkes, A. Green, J. M. F. Gunn, and B. J. Powell for helpful discussions.  Financial support from the EPSRC
(UK), SUPA,  CIFMC, FAPESP, and CNPq is gratefully acknowledged.  JQ
acknowledges an  Atlas Fellowship awarded by CCLRC (now STFC) in association
with St.\  Catherine's College, Oxford. VC acknowledges the pleasant
hospitality of the Instituto de F{\'\i}sica de S{\~a}o Carlos. CH is grateful
for the hospitality of the Centro de Giorgi in Pisa, where part of this work
was carried out.

* Email: j.quintanilla@rl.ac.uk (corresponding author)

\end{document}